\documentclass[final,1p,times]{elsarticle}

\usepackage{epsfig}

\usepackage{amssymb}

\journal{Journal of Magnetism and Magnetic Materials}

\begin{document}

\begin{frontmatter}



\title{Charge order, superconducting correlations, and $\mu$SR}


\author{J.E. Sonier}

\address{Department of Physics, Simon Fraser University, 8888 University Drive, Burnaby, V5A 1S6}

\begin{abstract}
The recent discoveries of short-range charge-density wave fluctuations in the normal state of several hole-doped cuprate superconductors constitute a 
significant addition to the known intrinsic properties of these materials. Besides likely being associated with the normal-state pseudogap, 
the charge-density wave order presumably influences the build-up of known superconducting correlations as the temperature is lowered toward the 
superconducting state. As a pure magnetic probe, muon spin rotation ($\mu$SR) is not directly sensitive to charge order, but may sense its presence
via the effect it has on the magnetic dipolar coupling of the muon with the host nuclei at zero or low magnetic field. At higher field where $\mu$SR is completely blind 
to the effects of charge order, experiments have revealed a universal inhomogeneous normal-state response extending to temperatures well above $T_c$.
The measured inhomogeneous line broadening has been attributed to regions of superconducting correlations that exhibit varying degrees of fluctuation diamagnetism. 
Here, the compatibility of these results with other measurements showing charge order correlations or superconducting fluctuations above $T_c$
is discussed.    
\end{abstract}


\end{frontmatter}



\section{Introduction}
\label{Intro}

Over the past year charge ordering has been established as a common feature of hole-doped high-temperature
cuprate superconductors. In different compounds and over a limited range of $p$ in the
underdoped regime (where $p$ is the number of doped holes per planar Cu atom), fluctuating charge order 
develops below a doping-dependent crossover temperature $T_{\rm CO}$ above the superconducting critical temperature $T_c$ 
\cite{Ghiringhelli,Chang,Torchinsky,Tacon,Comin,Neto}. In Bi$_2$Sr$_{2-x}$La$_x$CuO$_{6+\delta}$ (Bi-2201), $T_{\rm CO}$ has been observed to coincide 
with the pseudogap temperature $T^*$ \cite{Comin}, but is considerably below $T^*$ in other compounds. While the precise connection to the pseudogap 
is still being investigated, there is little doubt that the charge order competes with superconductivity.
X-ray scattering experiments show that the peak scattering intensity grows with decreasing temperature, but
decreases significantly below $T_c$ where the electrical resistivity vanishes \cite{Ghiringhelli,Chang,Neto}. Application of a magnetic field perpendicular 
to the CuO$_2$ layers has no effect above $T_c$, but enhances the scattering intensity below $T_c$. The latter can be understood as being due to the nucleation 
of charge order in and around the vortex-core regions, where the superconducting order parameter is suppressed. Evidence for this comes
from the detection of electronic modulations in the vortex cores of Bi$_2$Sr$_2$CaCu$_2$O$_{8+x}$ (Bi-2212) by scanning tunneling microscopy \cite{Hoffman} and
from nuclear magnetic resonance (NMR) measurements of the field-dependence of quadrupole splittings in the vortex-solid phase of 
YBa$_2$Cu$_3$O$_{6+x}$ (Y-123) 
\cite{Wu1}. In La$_{2-x-y}$Nd$_{y}$Sr$_x$CuO$_4$ (LNSCO) and 
La$_{2-x}$Ba$_x$CuO$_4$ (LBCO), where $T_c$ vanishes or is strongly suppressed near $p \! = \! 1/8$, the charge order freezes in zero field and 
coexists with static spin order, forming so-called ``stripes'' 
\cite{Tranquada,Fujita,Abbamonte}. Although static charge order in zero field has not been detected in other cuprates, an enhanced X-ray scattering intensity is 
observed in Y-123 and NdBa$_2$Cu$_3$O$_{6+x}$ near $p \! = \! 1/8$, where there is a plateau-like behavior of 
$T_c(p)$ \cite{Ghiringhelli}. In fact, it is in the vicinity of $p \! = \! 1/8$ where long-range static charge order is observed in Y-123 by NMR 
for high applied magnetic fields that locally \cite{Wu1} or fully suppress \cite{Wu2} superconductivity.

A notable feature of the charge order observed in cuprates under conditions where superconductivity is not fully suppressed, is that it is short range
above $T_c$. In Y-123 the in-plane correlation length is
doping dependent \cite{Ghiringhelli}, and increases with decreasing temperature up to a maximum of $\xi_{ab} \! \approx \! 10$~nm at $T_c$ for 
$p \! \approx \! 1/8$ \cite{Chang}. Moreover, the charge order is very weakly correlated along
the $c$-direction, with a correlation length $\xi_c \! \lesssim \! 1$ lattice units. In contrast, the charge order observed in Bi-2201 has a correlation 
length of $\xi_{ab} \! \approx \! 2-3$~nm, which evolves very little with temperature or doping \cite{Comin}. The correlation length of the charge order 
detected in underdoped Bi-2212 is also rather short \cite{Neto}. It has been suggested that nanometer-sized
regions of charge order are pinned by lattice defects, which are inherent in all superconducting cuprates \cite{Tacon}. The quasi-long-range static 
stripe order ($\xi_{ab} \! \approx \! 50$~nm) that emerges in LNSCO and LBCO near $p \! = \! 1/8$ is perhaps a dramatic demonstration of this, 
as its formation is concurrent with the onset of a low-temperature tetragonal (LTT) distortion, in which a change in the tilt axis of the CuO$_6$ octahedra 
creates structural anisotropy within the CuO$_2$ planes.  

Below $T_c$, the tendency toward charge order must compete with longe-range superconducting order. While the latter dominates, charge order may continue to
thrive near defects where the superconducting order parameter of the short-coherence length cuprates is naturally suppressed. In the vortex-solid state, there is a
natural spatially periodic suppression of the superconducting order parameter, although it is also energetically favorable for vortices to reside at the same locations 
as the defects. Above $T_c$ where there is no longer a competition with superconductivity, one may ask whether elucidating the regions in between the
short-range charge order is more relevant to understanding high-$T_c$ superconductivity? The low superconducting carrier
density of underdoped cuprates makes them susceptible to enhanced Cooper-pair phase fluctuations that ultimately destroy the superconducting state at $T_c$,
but Cooper pairs lacking long-range phase coherence can survive to higher temperatures \cite{Emery}. While there is ample experimental evidence
that this is indeed the case \cite{Corson,Xu,Wang:05,Wang:06,Rullier:06,Gomes,Cyr,Li,Dubroka,Grbic,Kondo,Bilbro}, 
there has been disagreement on the range of temperature over which superconducting correlations persist above $T_c$ and little
information on the spatial homogeneity of these correlations in the bulk.

Traces of superconductivity existing in a spatially inhomogeneous state is one way to reconcile the various conclusions that have emerged from 
experiments using different techniques. This picture is certainly not new. Superconducting ``droplets'' resulting from an intrinsic charge inhomogeneity
were previously invoked to explain an anomalous irreversible normal-state diamagnetism of Y-123 and La$_{2-x}$Sr$_x$CuO$_4$ (La-214), as measured by bulk 
magnetization \cite{Lanscia1,Lanscia2}. However, the only visual evidence of inhomogeneous superconducting correlations has come from scanning tunneling 
microscopy (STM) measurements on Bi-2212 (with $p \! \leq \! 0.22$), which revealed the nucleation of pairing gaps in nanometer-sized regions well 
above $T_c$ \cite{Gomes}. These regions proliferate as the temperature is lowered, with 100~\% gap coverage occurring near the surface of Bi-2212 when full-blown superconductivity takes place at $T_c$. The onset temperature $T_{\rm p}$ of the pairing-gap regions has a doping dependence that roughly follows $T_c$, but apparently peaks
somewhere in the underdoped regime (see Fig.~\ref{fig2}), where there is some difficulty in separating out the contribution of the pairing gap to the STM spectra from the 
contribution of the wider pseudogap. Infrared spectroscopy experiments \cite{Dubroka} that do not give spatially resolved information, indicate that $T_{\rm p}$ 
reaches a maximum value well below optimal doping (see Fig.~\ref{fig3}).

\section{$\mu$SR}
\label{muSR}

\subsection{Effect of charge order}
\label{charge}

The positive muon ($\mu^+$) used for $\mu$SR experiments on condensed matter systems has spin angular momentum of 1/2, and hence does not possess an electric 
quadrupole moment. Thus the $\mu^+$ is not directly sensitive to charge order, but rather is a pure magnetic probe that dipolar couples to the
nuclei of the host compound. Nuclei having spin greater than 1/2 (such as $^{63}$Cu and $^{65}$Cu) couple with the local electric field gradient (EFG), 
which is modified by the presence of the $\mu^+$, lattice structure changes and/or the development of charge order. The build up of charge-order correlations
has a temperature-dependent influence on the nuclear spins. On the other hand, the coupling of the nuclei to the EFG of the $\mu^+$ does 
not evolve with temperature; unless the muon diffuses during its short lifetime (The mean lifetime of the $\mu^+$ is $\tau_\mu \! \approx \! 2.2$~$\mu$s.). In cuprates, 
this apparently happens above $T \! \sim \! 200$~K (see for example Fig.~4 in Ref.~\cite{Sonier:02}). On the other hand, the build up of charge-order correlations
has a temperature-dependent influence on the magnetic-dipolar coupling of the host nuclei with the muon. 

Nuclear dipolar broadening of the internal magnetic field distribution $n(B)$ causes relaxation of the
$\mu$SR time-domain signal. The ``inhomogeneous'' line broadening due to the nuclear dipolar fields
may be probed by $\mu$SR in the absence of an applied magnetic field, or with a static field applied ``transverse'' 
to the initial muon spin polarization {\bf P}$(t \! = \! 0)$.
The dipolar broadening depends on the interaction of the nuclear electric quadrupole moment with the local EFG and the Zeeman interaction
of the nuclei with the applied static magnetic field {\bf H} \cite{Hartmann}. 
The time evolution of the transverse-field (TF) $\mu$SR polarization is  
\begin{equation}
P_{\rm TF}(t) = G(t) \cos(\omega_\mu t + \phi) \, ,
\end{equation}
where $\phi$ is the phase angle between the axis of the positron detector and {\bf P}$_{\rm TF}(t \! = \! 0)$, 
$\omega_{\mu}$ is the muon Larmor frequency and $G(t)$ is a function that describes the relaxation of the TF-$\mu$SR signal.  
In situations where the magnitude of the quadrupole couplings is larger or comparable 
to the Zeeman interactions, the quantization axes of the nuclear spins is dependent on both the direction of {\bf H} and the direction of the principal axes of the
EFG tensor \cite{Matthias}. However, only the Zeeman interactions depend on magnetic field, and at high magnetic field 
the nuclei are quantized in the direction of {\bf H}. In this so-called ``Van-Vleck'' limit \cite{VanVleck} the relaxation of the $\mu$SR signal due to 
magnetic dipolar interactions of the muon spin with the nuclear spins is related to the second moment of the local dipolar-field distribution, which depends 
only on the direction of the large field {\bf H}, and not its magnitude. For example, in pure Cu the Van-Vleck limit is reached above $H \! \sim \! 0.5$~T \cite{Camani}.
Likewise, any effect of charge order on the relaxation rate of the $\mu$SR signal for cuprates is most certainly absent for applied magnetic fields of order 1~T. 
Such is not the case for relaxation caused by a distribution of quasi-static or time-averaged internal magnetic fields from electronic magnetic moments, 
or from a distribution of time-averaged internal fields associated with inhomogeneous dynamic supercurrent screening of the applied magnetic field. 
These contributions to the TF-$\mu$SR relaxation rate are expected to exhibit significant and distinguishable dependences on the magnitude of {\bf H}.
\begin{figure}
\centering
\includegraphics[width=11cm]{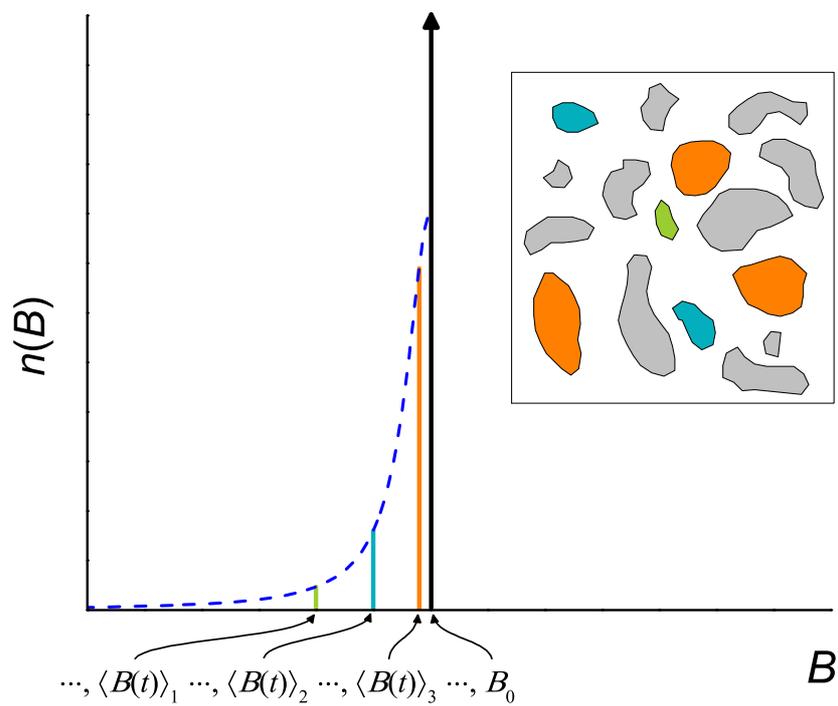}
\caption{(Color online) Schematic of the contribution to the probability distribution of magnetic field $n(B)$ experienced
by muons from random patches exhibiting different fluctuating diamagnetic responses; {\it i.e.} a ``left-half'' Lorentzian distribution 
of time-averaged fields (dashed blue curve). Broadening due to nuclear and electron magnetic dipolar fields is not included in the figure. In between 
the diamagnetic patches there is a uniform internal magnetic field $B_0$ that appears as a large spike in the probability distribution. Within each patch, 
the muons sense a time-averaged magnetic field $\langle B(t) \rangle_i$ less than $B_0$.}
\label{fig1}
\end{figure}
\subsection{High magnetic field}
\label{field}

In lightly-doped superconducting cuprates, the $\mu$SR signal at low temperatures and $H \! = \! 0$ is relaxed by quasi-static internal magnetic fields 
associated with remnant antiferromagnetic Cu-spin correlations of the undoped parent Mott insulating compound \cite{Kiefl,Niedermayer,Panagopoulos,Sanna}. 
While a magnetic field applied parallel to the $c$-axis results in a discernible relaxation from magnetism, which is extended to higher doping and higher 
temperature by the field, the effect is greatly diminished with increasing $p$ \cite{Savici}. 
Hence the effect of Cu magnetic moments on the relaxation rate of the TF-$\mu$SR signal at high magnetic field is easily recognizable. 
This is also true of the effect of paramagnetic moments that are present in heavily-overdoped cuprates. In an applied field the paramagnetism contributes a Curie-like temperature dependence to the TF-$\mu$SR relaxation rate and this contribution is enhanced with increasing $p$ \cite{MacDougall,Kaiser}. 
\begin{figure}
\centering
\includegraphics[width=13.0cm]{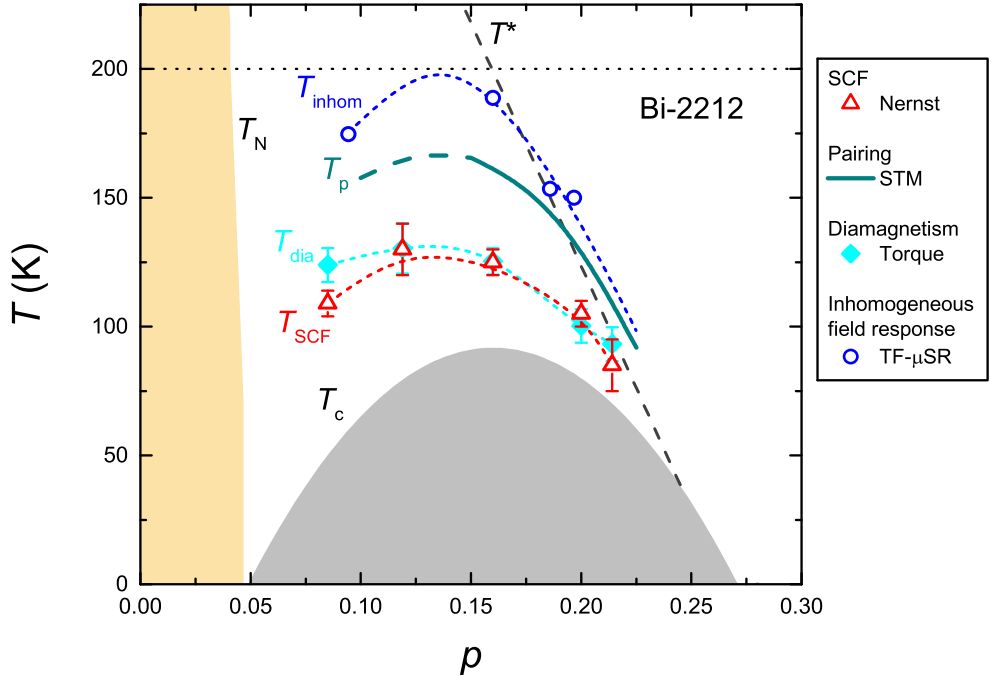}
\caption{(Color online) Onset temperatures for experimental signatures of superconducting correlations in Bi-2212. The data points for the onset of 
superconducting fluctuations (SCF) were obtained by vortex-Nernst effect measurements \cite{Wang:06}. The data points 
for $T_{\rm dia}$ signify the onset of a field-enhanced diamagnetic signal detected by torque magnetometry \cite{Wang:05}. The curve for $T_{\rm p}$
is from STM measurements \cite{Gomes}, and indicates the temperature above which less than 10~\% of the sample has a pairing gap. The dashed part of the
curve in the underdoped regime indicates the uncertainty in separating out the contribution to the STM spectra of the smaller pairing gap from that of the 
wider pseudogap. The data points for $T_{\rm inhom}$ indicate the temperature above which the exponential TF-$\mu$SR relaxation rate $\Lambda$ at $H \! = \! 7$~T is 
less than 0.02~$\mu$s$^{-1}$ \cite{Lotfi}. The short dashed curves through the data points are guides to the eye. The dotted horizontal line indicates the temperature
above which muon diffusion may occur. The pseudogap temperature $T^*$ is inferred from tunneling spectroscopy measurements \cite{Ren}.}
\label{fig2}
\end{figure}   
Distinct from the effects of magnetism is a weak field-induced exponential contribution to the relaxation of the normal-state TF-$\mu$SR signal that roughly 
follows $T_c$. This was first identified in Y-123 and La$_{2-x}$Sr$_x$CuO$_4$ (La-214) \cite{Sonier:08}, and more recently observed in Bi-2212 \cite{Lotfi}.
This normal-state exponential relaxation rate (denoted by $\Lambda$) appears to originate from inhomogeneous superconducting correlations. Among the
evidence for this is a suppression of $\Lambda$ in Y-123 near $p \! = \! 1/8$ \cite{Sonier:08}, and a universal scaling of $\Lambda$ with the maximum value of $T_c$ 
for each cuprate family \cite{Lotfi}. Further support for this interpretation is the simlarity of the doping dependences of $\Lambda$ and 
the onset temperatures above $T_c$ for Cooper pairing or superconducting fluctuations identified by other techniques (see Figs.~\ref{fig2} and \ref{fig3}). 

\subsection{Effect of diamagnetic regions}
\label{Diamagnetic}

\begin{figure}
\centering
\includegraphics[width=13.0cm]{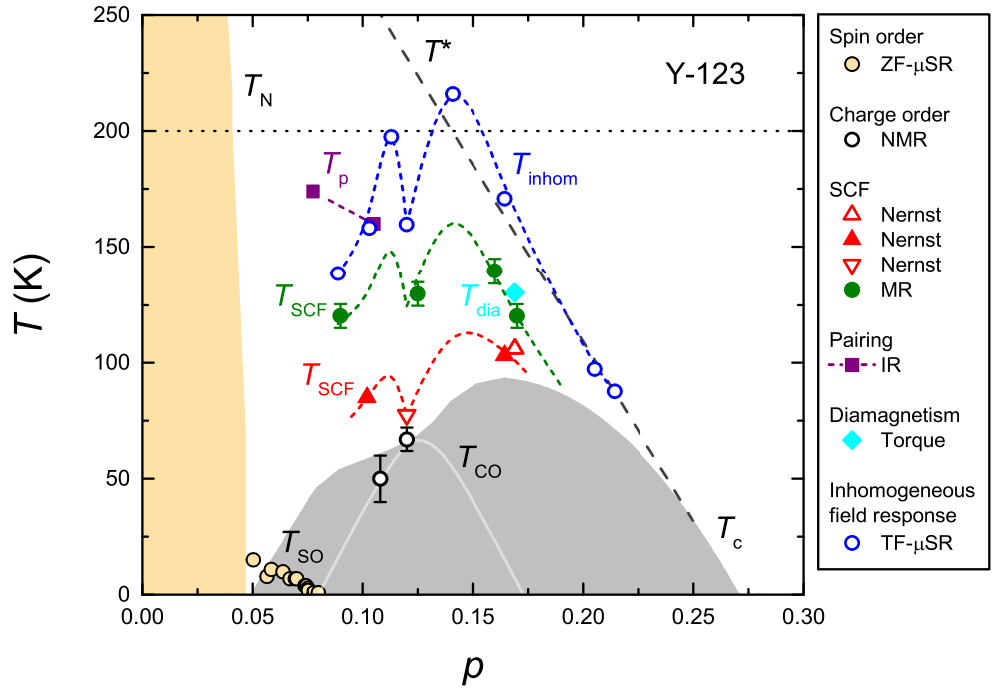}
\caption{(Color online) Onset temperatures for experimental signatures of superconducting correlations in Y-123. 
The data points for the onset of short-range static spin order (SO) come from zero-field (ZF) $\mu$SR measurements, and the onset of static charge order (CO) from
high-field NMR experiments \cite{Wu2}. The data points for $T_{\rm inhom}$ indicate the temperature 
above which the exponential TF-$\mu$SR relaxation rate $\Lambda$ at $H \! = \! 7$~T is less than 0.02~$\mu$s$^{-1}$ \cite{Sonier:08,Lotfi}. 
Note that the two data points above $p \! = \! 0.20$ are Ca-doped Y-123. The short-dashed curve through 
these data points is a guide to the eye. The dotted horizontal line indicates the temperature above which muon diffusion likely occurs. 
The red data points for the onset of superconducting fluctuations (SCF) were obtained by vortex-Nernst effect 
($\vartriangle$ \cite{Wang:06}, $\blacktriangle$ \cite{Rullier:06}, and $\nabla$ \cite{Daou}) and magnetoresistance (MR) 
measurements \cite{Rullier:11}. The short-dashed curves through these data points show that the hole-doping dependence of existing data 
for $T_{\rm SCF}$ is compatible with the $\mu$SR measurements of $\Lambda$. The data point denoted by $T_{\rm dia}$ signifies the onset of a 
field-enhanced diamagnetic signal detected by torque magnetometry \cite{Li}. The data points for $T_{\rm p}$ were obtained by infrared (IR) spectrosopy. 
The pseudogap temperature $T^*$ is inferred from Nernst effect and resistivity measurements \cite{Daou}.}
\label{fig3}
\end{figure}

As explained in Ref.~\cite{Lotfi}, the motion of vortices in a uniform liquid state is too fast to cause the observed inhomogeneous line broadening above $T_c$.
Sensitivity of $\mu$SR to superconducting fluctuations requires regions with varying degrees of the associated fluctuation diamagnetism. 
In this case, the implanted muons experience distinct time-averaged local fields $\langle B(t) \rangle_i$ ($i \! = \! 1$, 2, 3, ...) in diamagnetic regions
with different degrees of supercurrent screening. 
Figure~\ref{fig1} shows a schematic of the probability distribution of magnetic field for muons probing irregular-shaped diamagnetic patches of varying phase stiffness 
within an otherwise normal sample. The figure assumes a ``left-half'' Lorentzian distribution 
of time-averaged fields (dashed blue curve), as this introduces a corresponding exponential relaxation $\exp(-at)$ 
of $P_{\rm TF}(t)$, where $a$ is a relaxation rate.
The essentially static nuclear magnetic dipolar fields, as well as dipolar fields from 
fluctuating electron magnetic moments, will each further broaden the field distribution shown in Fig.~\ref{fig1}. 
The corresponding muon spin polarization function is     
\begin{equation}
P_{\rm TF}(t) = G_{\rm nuc}(t)G_{\rm el}(t)[f \exp(-a t)\cos(\gamma_\mu \overline{\langle B(t) \rangle} t) + (1-f)\cos(\gamma_\mu B_0 t)] \, ,
\label{eq:PolSum}
\end{equation}
where $G_{\rm nuc}(t)$ is a {\it temperature-independent} function accounting for relaxation due to the nuclear magnetic dipolar fields, 
$G_{\rm el}(t)$ is a {\it temperature-dependent} function accounting for relaxation due to electron magnetic dipolar fields, 
$f$ is the volume fraction of the sample exhibiting diamagnetism, $\overline{\langle B(t) \rangle}$ is the mean value of all the time-averaged 
local fields associated with the diamagnetic regions, and $B_0$ is the average field between them. Since the diamagnetic shift in field for each 
region is extremely small compared to the applied field, then $\overline{\langle B(t) \rangle} \! \approx \! B_0$. Furthermore, the width of 
the distribution of weak diamagnetic
fields and the corresponding relaxation rate $a$ are small, so that $\exp(-at) \! \approx \! 1 - a t$. 
Hence Eq.~(\ref{eq:PolSum}) may be approximated as follows 
\begin{eqnarray}
P_{\rm TF}(t) & \approx & G_{\rm nuc}(t) G_{\rm el}(t) [f (1-a t)\cos(\gamma_\mu B_0 t) + (1-f)\cos(\gamma_\mu B_0 t)]  \nonumber \\
              & = & G_{\rm nuc}(t) G_{\rm el}(t) (1-fat)\cos(\gamma_\mu B_0 t) \nonumber \\
              & \approx & G_{\rm nuc}(t) G_{\rm el}(t) \exp(-\Lambda t)\cos(\gamma_\mu B_0 t) \, ,   
\label{eq:Approx}
\end{eqnarray}
where $\Lambda = fa$.         
For rapidly fluctuating electron dipolar magnetic fields, $G_{\rm el}(t) \exp(-\lambda_{\rm el}t)$. Consequently, contributions
from fast-fluctuating electronic spins must be determined before ascribing any exponential-relaxation component to 
fluctuation diamagnetism. With this said, for the range
of hole-doping considered in Refs.~\cite{Sonier:08,Lotfi}, the electronic spin fluctuations above $T_c$ are
too fast for the muons to follow. In particular, the fluctuating electron dipolar magnetic fields sensed by the muon are
so fast that they average to zero over a period of time that is considerably shorter than the muon lifetime, so that $\lambda_{\rm el} \! \approx \! 0$. 
This presumption is first based on the limited ranges of $p$ and $T$ that static and fluctuating electronic magnetic 
moments have been detected by $\mu$SR at zero field \cite{Kiefl,Niedermayer,Panagopoulos,Sanna}. The observed doping dependence 
of $\Lambda(T \! > \! T_c)$ at $H \! = \! 7$~T is also incompatible with the exponential relaxation being dominated by 
electron dipolar magnetic fields. Specifically, $\Lambda(T \! > \! T_c)$ within the underdoped regime decreases as $p$ is lowered toward 
the Mott insulator. Furthermore, $\Lambda(T \! > \! T_c)$ dips in Y-123 near $p \! = \! 1/8$, where $T_c$ plateaus \cite{Sonier:08,Lotfi} and charge
order is most pronounced \cite{Ghiringhelli,Wu2} (see Fig.~\ref{fig3}). The latter is a clear indication of the insensitivity
of the $\mu^+$ to the effects of charge order at high magnetic field. 
   
\section{Perspective on relationship to other experiments}

Much of the evidence for superconducting correlations persisting above $T_c$ has come from methods that rely on what is essentially
a bulk response, incapable of distinguishing between spatially homogeneous and inhomogeneous systems. 
The spatially-resolved STM measurements of Bi-2212 by Gomes {\it et al.} \cite{Gomes} indicate that the percentage of the sample occupied 
by random nanometer-sized pairing gap regions decreases as the temperature is elevated further above $T_c$.  
Assuming the STM gap maps on Bi-2212 are indicative of the bulk and the observed pairing regions sustain short-range phase coherence, some bulk signals
indicative of superconducting correlations will vanish or weaken when Josephson tunneling between the regions is lost or these regions 
become fairly dilute.
Indeed the onset temperatures for the vortex-Nernst and field-enhanced diamagnetic signals of Bi-2212 lie well below the temperature at which STM
detects pairing gaps in approximately 10~\% of the sample (see Fig.~\ref{fig2}). On the other hand, muons sample random locations in
this patchy diamagnetic environment and provided there is a variation in the degree of diamagnetic supercurrent screening by these
diminishing regions, will continue to sense a distribution of time-averaged magnetically inequivalent sites. 
Hence the corresponding relaxation of the TF-$\mu$SR signal may persist to higher temperatures than signatures of superconducting
correlations detected by other methods. 

It is important to stress that the characteristic time window of fluctuating local magnetic fields that can be sensed by $\mu$SR demands that 
superconducting fluctuations above $T_c$ be confined to isolated or weakly connected regions. Note that quantitative
justification for this statement is given in Ref.~\cite{Lotfi}. Consequently, the TF-$\mu$SR
measurements essentially extend the information provided by the STM gap maps on Bi-2212. First, they do so by verifying the evolution
of the spatially-varying pairing correlations in the underdoped regime. While a pairing gap above $T_c$ is clearly identified in the
STM spectra of overdoped Bi-2212, in the underdoped regime only a low-energy kink within the wider pseudogap is visible. 
Yet as shown in Fig.~\ref{fig2}, the doping dependence of the inhomogeneous magnetic response detected by TF-$\mu$SR 
follows the pairing-gap coverage that is less accurately determined by STM in the underdoped regime. 
Second, the combined TF-$\mu$SR results on Bi-2212, Y-123 and La-214 \cite{Sonier:08,Lotfi} 
imply that inhomogeneous superconducting correlations above $T_c$ are not limited to the surface of Bi-2212, but rather are a universal bulk property of 
hole-doped cuprates. Furthermore, because the degree of chemical disorder is very different in these three compounds, 
the electronic inhomogeneity must be deemed an intrinsic property.  

It is clear that the experimentally determined onset temperature ($T_{\rm onset}$) for superconducting correlations is very much dependent on the
sensitivity of the method. Moreover, the build up of superconducting correlations in different regions and the reduction in phase fluctuations of the 
superconducing order parameter is a gradual process, with the TF-$\mu$SR relaxation rate $\Lambda$ \cite{Sonier:08,Lotfi}, the vortex-Nernst 
signal \cite{Wang:06}, and the diamagnetic signal \cite{Li} all slowly decaying with increased temperature above $T_c$.    
The TF-$\mu$SR signal is ultimately influenced by significant muon diffusion above $T \! \sim \! 200$~K, which prevents an accurate determination of
$T_{\rm onset}$ at certain hole dopings. Nevertheless, while different techniques give different values of $T_{\rm onset}$, the varied results
are consistent in showing that $T_{\rm onset}$ and the pseudogap temperature $T^*$ strongly diverge with decreasing $p$ in the underdoped 
regime. This and the observation of distinct pseudo and superconducting (or pairing) gaps by various methods \cite{Hufner} strongly suggests that the pseudogap 
is not associated with preformed pairs, but instead is associated with a generic electronic phase separation that leads to charge ordering.
While the TF-$\mu$SR measurements do not contribute any information on the charge order, they support a picture in which the pseudogap regime is 
predominantly a mixture of at least two distinct regions, with one supporting charge-order fluctuations and the other sustaining superconducting fluctuations.\\          

\noindent {\bf Acknowledgements}
This work is supported in part by NSERC and the Canadian Institute for Advanced Research.


\begin{thebibliography}{00}


\bibitem{Ghiringhelli} G. Ghiringhelli, M. Le Tacon, M. Minola, S. Blanco-Canosa, C. Mazzoli, N.B. Brookes, G.M. De Luca, A. Frano, D.G. Hawthorn, F. He, T. Loew, M. Moretti Sala, D.C. Peets, M. Salluzzo, R. Sutarto, G.A. Sawatzky, E. Weschke, B. Keimer, L. Braicovich, Science 337 (2012) 821.  

\bibitem{Chang} J. Chang, E. Blackburn, A.T. Holmes, N.B. Christensen, J. Larsen, J. Mesot, R. Liang, 
D.A. Bonn, W.N. Hardy, A. Watenphul, M.V. Zimmermann, E.M. Forgan, and S.M. Hayden, Nature Physics 8 (2012) 871.

\bibitem{Torchinsky} D.H. Torchinsky, F. Mahmood, A.T. Bollinger, I. Bo\u{z}ovi\'{c}, N. Gedik,
Nature materials 12 (2013) 387.

\bibitem{Tacon} M. Le Tacon, A. Bosak, S.M. Souliou, G. Dellea, T. Loew, R. Heid, K.-P. Bohnen, 
G. Ghiringhelli, M. Krisch, B. Keimer, Nature Physics 10 (2014) 52.

\bibitem{Comin} R. Comin, A. Frano, M.M. Yee, Y. Yoshida, H. Eisaki, E. Schierle, E. Weschke, R. Sutarto, 
F. He, A. Soumyanarayanan, Y. He, M. Le Tacon, I.S. Elfimov, J.E. Hoffman, G.A. Sawatzky, B. Keimer, A. Damascelli, Science 343 (2014) 390.

\bibitem{Neto} E.H. da Silva Neto, P. Aynajian, A. Frano, R. Comin, E. Schierle, E. Weschke, 
A. Gyenis, J. Wen, J. Schneeloch, Z. Xu, S. Ono, G. Gu, M. Le Tacon, A. Yazdani, Science 343 (2014) 393.

\bibitem{Hoffman} J.E. Hoffman, E.W. Hudson, K.M. Lang, V. Madhavan, H. Eisaki, S. Uchida, J.C. Davis, Science 295 (2002) 466.

\bibitem{Wu1} T. Wu, H. Mayaffre, S. Kr\"{a}mer, M. Horvati\'{c}, C. Berthier, P.L. Kuhns, A.P. Reyes, R. Liang, W.N. Hardy, D.A. Bonn, M.H. Julien, 
Nature Communications 4 (2013) 2113.

\bibitem{Wu2} T. Wu, H. Mayaffre, S. Kr\"{a}mer, M. Horvati\'{c}, C. Berthier, W.N. Hardy, R. Liang, D.A. Bonn, M.H. Julien, Nature 477 (2011) 191.

\bibitem{Tranquada} J.M. Tranquada, B.J. Sternlieb, J.D. Axe, Y. Nakamura, S. Uchida, Nature 375 
(1995) 561.

\bibitem{Fujita} M. Fujita, H. Goka, K. Yamada, M. Matsuda, Phys. Rev. Lett. 88 (2002) 167008. 

\bibitem{Abbamonte} P. Abbamonte, A. Rusydi, S. Smadici, G.D. Gu, G.A. Sawatzky, D.L. Feng,
Nature Physics 1 (2005) 155.

\bibitem{Emery} V.J. Emery, S.A. Kivelson, Nature 374 (1995) 434. 

\bibitem{Corson} J. Corson, R. Mallozzi, J. Orenstein, J.N. Eckstein, I.~Bozovic, Nature 398 (1999) 221.

\bibitem{Xu} Z.A. Xu, N.P. Ong, Y. Wang, T. Kakeshita, S. Uchida, Nature 406 (2000) 486.

\bibitem{Wang:05} Y. Wang, L. Li, M.J. Naughton, G.D. Gu, S. Uchida, N.P. Ong, Phys. Rev. Lett. 95 (2005) 247002.

\bibitem{Wang:06} Y. Wang, L. Li, N.P. Ong, Phys. Rev. B 73 (2006) 024510.

\bibitem{Rullier:06} F. Rullier-Albenque, R. Tourbot, H. Alloul, P. Lejay, D. Colson, A. Forget, Phys. Rev. Lett. 96 (2006) 067002.  

\bibitem{Gomes} K.K. Gomes, A.N. Pasupathy, A. Pushp, S. Ono, Y. Ando, A. Yazdani, Nature 447 (2007) 569.

\bibitem{Cyr} O. Cyr-Choini\`{e}re, R. Daou, F. Lalibert\'{e}, D. LeBoeuf, N. Doiron-Leyraud, J. Chang, J.-Q. Yan, J.-G. Cheng, J.-S. Zhou, J.B. Goodenough,
S. Pyon, T. Takayama, H. Takagi, Y. Tanaka, L. Taillefer, Nature 458 (2009) 743.

\bibitem{Daou} R. Daou, J. Chang, D. LeBoeuf, O. Cyr-Choini\`{e}re, F. Lalibert\'{e}, N. Doiron-Leyraud, B.J. Ramshaw,
R. Liang, D.A. Bonn, W.N. Hardy, L. Taillefer, Nature 463 (2010) 519.
 
\bibitem{Li} L. Li, Y. Wang, S. Komiya, S. Ono, Y. Ando, G.D. Gu, N.P. Ong, Phys. Rev. B 81 (2010) 054510. 

\bibitem{Dubroka} A. Dubroka, M. R\"{o}ssle, K.W. Kim, V.K. Malik, D. Munzar, D.N. Basov, A.A. Schafgans, S.J. Moon, C.T. Lin, D. Haug,
V. Hinkov, B. Keimer, Th. Wolf, J.G. Storey, J.L. Tallon, C. Bernhard, Phys. Rev. Lett. 106 (2011) 047006.

\bibitem{Grbic} M. S. Grbi\'{c}, M. Po\v{z}ek, D. Paar, V. Hinkov, M. Raichle, D. Haug, B. Keimer, N. Bari\v{s}i\'{c}, A. Dul\v{c}i\'{c}, Phys. Rev. B 83 (2011) 144508.

\bibitem{Kondo} T. Kondo, Y. Hamaya, A.D. Palczewski, T. Takeuchi, J.S. Wen, Z.J. Xu, G. Gu, J. Schmalian, A. Kaminski, Nature Physics 7 (2011) 21. 

\bibitem{Bilbro} L.S. Bilbro, R. Vald\'{e}s Aguilar, G. Logvenov, O. Pelleg, I. Bo\v{z}ovi\'{c}, N.P. Armitage, Nature Physics 7 (2011) 298.

\bibitem{Rullier:11} F. Rullier-Albenque, H. Alloul, G. Rikken, Phys. Rev. B 84 (2011) 014522.

\bibitem{Lanscia1} A. Lascialfari, A. Rigamonti, L. Romano, P. Tedesco, A. Varlamov, D. Embriaco, Phys. Rev. B 65 (2002) 144523.

\bibitem{Lanscia2} A. Lascialfari, A. Rigamonti, L. Romano, A. Varlamov, I. Zucca, Phys. Rev. B 68 (2003) 100505.

\bibitem{Sonier:02} J.E. Sonier, J.H. Brewer, R.F. Kiefl, R.H. Heffner, K.F. Poon, S.L. Stubbs, G.D. Morris, R.I. Miller, W.N. Hardy, R. Liang,
D.A. Bonn, J.S. Gardner, C.E. Stronach, N.J. Curro, Phys. Rev. B 66 (2002) 134501. 

\bibitem{Hartmann} O. Hartmann, Phys. Rev. Lett. 39 (1977) 832.

\bibitem{Matthias} E. Matthias, W. Schneider, R.M. Steffen, Phys. Rev. 125 (1962) 261.

\bibitem{VanVleck} J.H. Van Vleck, Phys. Rev. 74 (1948) 1168.

\bibitem{Camani} M. Camani, F.N. Gygax, W. R\"{u}egg, A. Schenck, H. Schilling, Phys. Rev. Lett. 39 (1977) 836.

\bibitem{Kiefl} R.F. Kiefl, J.H. Brewer, J. Carolan, P. Dosanjh, W.N. Hardy, R. Kadono, J.R. Kempton, R. Krahn, P. Schleger, B.X. Yang, H. Zhou, G.M. Luke,
B. Sternlieb, Y.J. Uemura, W.J. Kossler, X.H. Yu, E.J. Ansaldo, H. Takagi, S. Uchida, C.L. Seaman, Phys. Rev. Lett. 63 (1989) 2136.

\bibitem{Niedermayer} Ch. Niedermayer, C. Bernhard, T. Blasius, A. Golnik, A. Moodenbaugh, J.I. Budnick, Phys. Rev. Lett. 80 (1998) 3843.

\bibitem{Panagopoulos} C. Panagopoulos, J.L. Tallon, B.D. Rainford, T. Xiang, J.R. Cooper, C.A. Scott, Phys. Rev. Lett. 66 (2002) 064501.

\bibitem{Sanna} S. Sanna, G. Allodi, G. Concas, A.D. Hillier, R. De Renzi, Phys.~Rev.~Lett. 93 (2004) 207001.

\bibitem{Savici} A.T. Savici, A. Fukaya, I.M. Gat-Malureanu, T. Ito, P.L. Russo, Y.J. Uemura, C.R. Wiebe, P.P. Kyriakou,
G.J. MacDougall, M.T. Rovers, G.M. Luke, K.M. Kojima, M. Goto, S. Uchida, R. Kadono, K. Yamada, S. Tajima, T. Masui, H. Eisaki,
N. Kaneko, M. Greven, G.D. Gu, Phys. Rev. Lett. 95 (2005) 157001.

\bibitem{MacDougall} G.J. MacDougall, A.T. Savici, A.A. Aczel, R.J. Birgeneau, H. Kim, S.-J. Kim, T. Ito, 
J.A. Rodriguez, P.L. Russo, Y.J. Uemura, S. Wakimoto, C.R. Wiebe, G.M. Luke, Phys. Rev. B 81 (2010) 014508.

\bibitem{Kaiser} C.V. Kaiser, W. Huang, S. Komiya, N.E. Hussey, T. Adachi, Y. Tanabe, Y. Koike, J.E. Sonier, Phys. Rev. B 86 (2012) 054522.

\bibitem{Sonier:08} J.E. Sonier, M. Ilton, V. Pacradouni, C.V. Kaiser, S.A. Sabok-Sayr, Y. Ando, S. Komiya,
W.N. Hardy, D.A. Bonn, R. Liang, and W.A. Atkinson, Phys. Rev. Lett. 101 (2008) 117001.

\bibitem{Lotfi} Z. Lotfi Mahyari, A. Cannell, E.V.L. de Mello, M. Ishikado, H. Eisaki, R. Liang, D.A. Bonn, J.E. Sonier, Phys. Rev. B 88 (2013) 144504.

\bibitem{Ren} J.K. Ren, X.B. Zhu, H.F. Yu, Y. Tian, H.F. Yang, C.Z. Gu, N.L. Wang, Y.F. Ren, S.P. Zhao, Scientific Reports 2 (2012) 248.

\bibitem{Hufner} S. H\"{u}fner, M.A. Hossain, A. Damascelli, G.A. Sawatzky, Rep. Prog. Phys. 71 (2008) 062501.

\end{thebibliography}
\end{document}